\documentclass[amssymb,prb,twocolumn,superscriptaddress,floats,showkeys,amsmath]{revtex4-1}

\usepackage{graphicx}
\usepackage{epstopdf}
\usepackage{hyperref} 
\hypersetup{colorlinks,citecolor=blue, filecolor=blue ,linkcolor=blue , urlcolor=blue, pdftex}

\usepackage[labeled,resetlabels]{multibib}

\newcites{SI}{SupI}

\begin{document}

\title{Probing and manipulating valley coherence of dark excitons in monolayer WSe$_2$}

\author{M. R. Molas}
\email{maciej.molas@fuw.edu.pl}
\affiliation{Laboratoire National des Champs Magn\'etiques Intenses, CNRS-UGA-UPS-INSA-EMFL, 25 avenue des Martyrs, 38042 Grenoble, France}
\affiliation{Institute of Experimental Physics, Faculty of Physics, University of Warsaw, ul. Pasteura 5, 02-093 Warszawa, Poland}
\author{A. O.~Slobodeniuk}
\affiliation{Laboratoire National des Champs Magn\'etiques Intenses, CNRS-UGA-UPS-INSA-EMFL, 25 avenue des Martyrs, 38042 Grenoble, France}
\author{T.~Kazimierczuk}
\affiliation{Institute of Experimental Physics, Faculty of Physics, University of Warsaw, ul. Pasteura 5, 02-093 Warszawa, Poland}
\author{K.~Nogajewski}
\affiliation{Laboratoire National des Champs Magn\'etiques Intenses, CNRS-UGA-UPS-INSA-EMFL, 25 avenue des Martyrs, 38042 Grenoble, France}
\affiliation{Institute of Experimental Physics, Faculty of Physics, University of Warsaw, ul. Pasteura 5, 02-093 Warszawa, Poland}
\author{M.~Bartos}
\affiliation{Laboratoire National des Champs Magn\'etiques Intenses, CNRS-UGA-UPS-INSA-EMFL, 25 avenue des Martyrs, 38042 Grenoble, France}
\author{P.~Kapu\'{s}ci\'{n}ski}
\affiliation{Laboratoire National des Champs Magn\'etiques Intenses, CNRS-UGA-UPS-INSA-EMFL, 25 avenue des Martyrs, 38042 Grenoble, France}
\affiliation{Department of Experimental Physics, Faculty of Fundamental Problems of Technology, Wroc{\l}aw University of Science and Technology, 27 Wybrze\.{z}e Wyspia\'{n}skiego, 50-370 Wroc{\l}aw, Poland}
\author{K.~Oreszczuk}
\affiliation{Institute of Experimental Physics, Faculty of Physics, University of Warsaw, ul. Pasteura 5, 02-093 Warszawa, Poland}
\author{K. Watanabe}
\affiliation{National Institute for Materials Science, 1-1 Namiki, Tsukuba 305-0044, Japan}
\author{T. Taniguchi}
\affiliation{National Institute for Materials Science, 1-1 Namiki, Tsukuba 305-0044, Japan}
\author{C. Faugeras}
\affiliation{Laboratoire National des Champs Magn\'etiques Intenses, CNRS-UGA-UPS-INSA-EMFL, 25 avenue des Martyrs, 38042 Grenoble, France}
\author{P. Kossacki}
\affiliation{Institute of Experimental Physics, Faculty of Physics, University of Warsaw, ul. Pasteura 5, 02-093 Warszawa, Poland}
\author{D. M. Basko}
\affiliation{Laboratoire de Physique et Mod\'elisation des Milieux Condens\'es, Universit\'e Grenoble Alpes and CNRS, 25 avenue des Martyrs, 38042 Grenoble, France}
\author{M. Potemski}
\email{marek.potemski@lncmi.cnrs.fr}
\affiliation{Laboratoire National des Champs Magn\'etiques Intenses, CNRS-UGA-UPS-INSA-EMFL, 25 avenue des Martyrs, 38042 Grenoble, France}
\affiliation{Institute of Experimental Physics, Faculty of Physics, University of Warsaw, ul. Pasteura 5, 02-093 Warszawa, Poland}

\begin{abstract}
	
Monolayers of semiconducting transition metal dichalcogenides are two-dimensional direct-gap systems which host tightly-bound excitons with an internal degree of freedom corresponding to the valley of the constituting carriers. Strong spin-orbit interaction and the resulting ordering of the spin-split subbands in the valence and conduction bands makes the lowest-lying excitons in WX$_2$ (X~being S or Se) spin-forbidden and optically dark. With polarization-resolved photoluminescence experiments performed on a WSe$_2$ monolayer encapsulated in a hexagonal boron nitride, we show how the intrinsic exchange interaction in combination with the applied in-plane and/or \mbox{out-of-plane} magnetic fields enables one to probe and manipulate the valley degree of freedom of the dark excitons.

\end{abstract}

\maketitle

Monolayers of transition metal dichalcogenides (TMDs), such as MX$_2$  with M=Mo or W, and X=S, Se or Te, are two-dimensional
direct-gap semiconductors~\cite{Mak2010} which attract a lot of interest due to their unique physical properties and potential
applications in optoelectronics, photonics and the development of valleytronics~\cite{Wang2012, Eda2013, Xia2014, Xu2014,
	Koperski2017, Wang2018}. The direct bandgap in semiconducting TMDs (S-TMDs) is located at the two inequivalent $\mathbf{K}_\pm$ 
points (valleys) of the first Brillouin zone, related by time reversal symmetry. In monolayers, the tightly bound and optically 
bright excitons~\cite{Chernikov2014, Ye2014, Ugeda2014, He2014} from $\mathbf{K}_\pm$ valley can efficiently couple to light with
right/left circular polarization\cite{Yao2008, Xiao2012}, respectively.

\begin{figure}
	\includegraphics[width=0.47\textwidth]{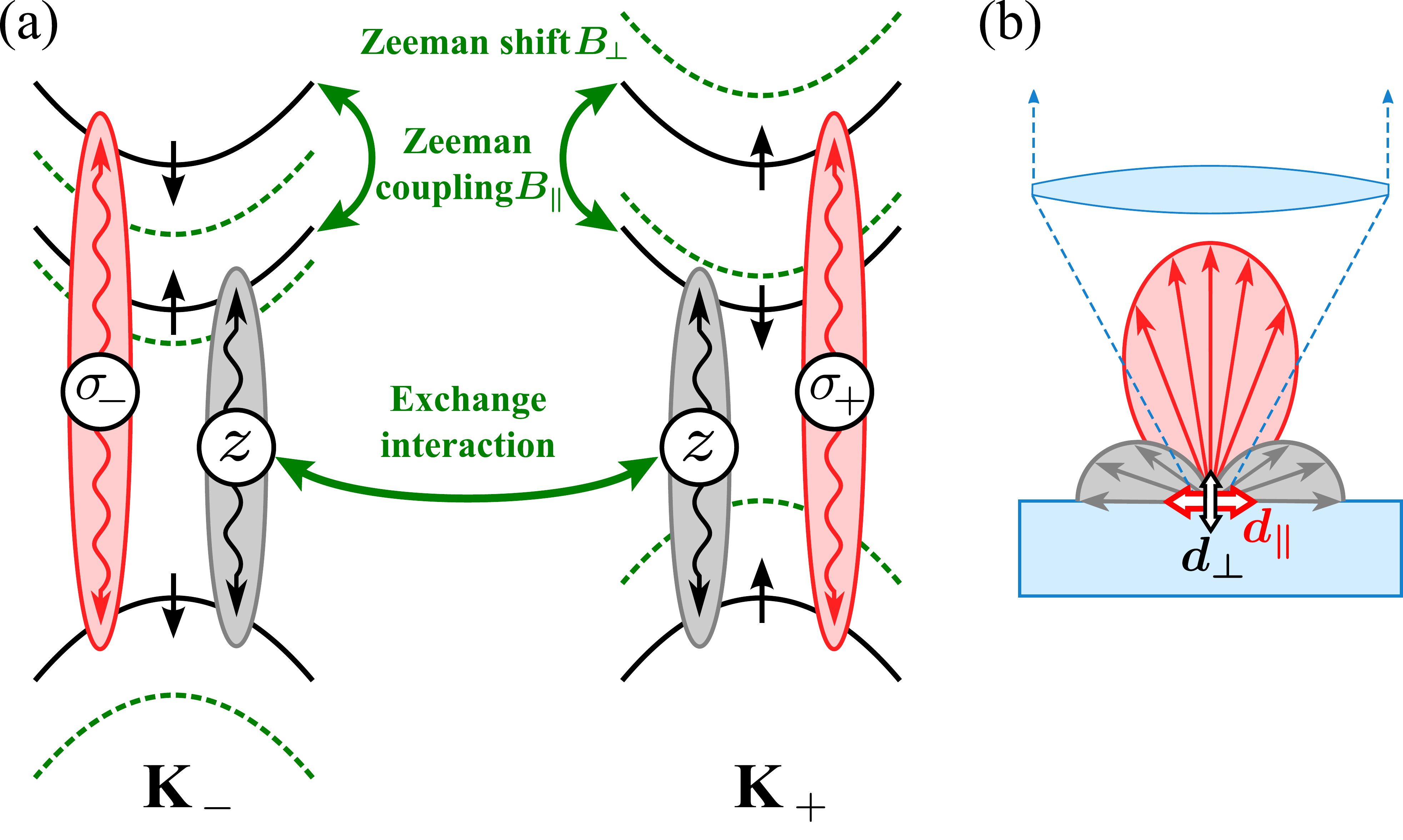}
	\caption{(a) Spin and valley structure of WSe$_2$. Black solid curves show the spin-split bands in the $\mathbf{K}_\pm$ valleys at zero magnetic field, the spin projections are indicated by black arrows. Grey and red ellipses represent spin-forbidden and spin-allowed excitonic transitions in each valley, respectively. Dashed green curves show Zeeman-shifted bands in a perpendicular magnetic field. Thick green arrows represent the exchange interaction which mixes the two valleys, and the Zeeman coupling in an in-plane magnetic field which mixes different spin subbands in the same valley. (b)~A~schematic representation of radiation patterns of dark and bright excitons in S-TMD monolayer at zero magnetic field. The grey (red) arrows indicate the propagation direction of the light, emitted by dark (bright) excitons. The blue dashed lines define the region of light collected by the lens. The black and red double-headed arrows depict the relative magnitudes and directions of out-of-plane ($d_\perp$) and in-plane ($d_\|$) dipoles.} \label{fig:excitons}
\end{figure}

A unique feature of S-TMD monolayers is the so-called spin-valley locking~\cite{Xiao2012}: strong spin-orbit interaction lifts the
degeneracy between the two spin projections $s=\uparrow,\downarrow$ in each valley, leaving only the Kramers degeneracy between 
the opposite valleys, $\mathbf{K}_+,s\leftrightarrow\mathbf{K}_-,-s$. While the valence band spin-orbit splitting $\Delta_v$ is 
very large (several hundred meV~\cite{Xiao2012}), its conduction band counterpart $\Delta_c$ is by an order of magnitude 
smaller~\cite{Kosmider2013, Kormanyos2015, Echeverry2016, Molas2017, Zhang2017}, thus allowing some degree of manipulation by an
in-plane magnetic field. In tungsten-based S-TMDs, $\Delta_c$ has the same sign as $\Delta_v$, leading to the spin subband ordering 
shown in Fig.~\ref{fig:excitons}(a). In each valley, the optically bright exciton has a higher energy than the dark exciton which 
is composed of the conduction and valence electronic states with opposite spin projections. This, among other things, results in 
a strong temperature dependence of the photoluminescence (PL) efficiency, as the bright states become more populated at higher
temperatures~\cite{Arora2015, Zhang2015, Wang2015}.

Dark excitons can couple to light only via a residual spin-flip dipole matrix element~$d_\perp$, whose direction is perpendicular
to the monolayer plane~\cite{Glazov2014, Slobodeniuk2016, WangMarie2017}. In consequence, the emission of dark excitons is directed 
predominantly along the monolayer plane, in contrast to the out-of-plane emission of bright excitons characterized by strong in-plane
optical dipoles, $d_\|$, see Fig.~\ref{fig:excitons}(b). Notably, the valley degeneracy of dark excitons is lifted by the exchange 
interaction which mixes the valleys and produces two eigenstates with different energies. The higher energy component takes up the 
whole oscillator strength due to~$d_\perp$ (so we call it ``grey''), while the low-energy component is completely decoupled from light 
and truly ``dark''~\cite{Dery2015,Slobodeniuk2016}.

The possibility to control the valley degree of freedom for bright excitons by optical polarization~\cite{Zeng2012, Mak2012, Jones2013} 
and external magnetic field~\cite{Aivazian2015, Wang2016, Smolenski2016} is impaired by the quick valley relaxation due to the exchange 
interaction~\cite{Glazov2014, Hao2016}, because the latter (i)~depends on the exciton center-of-mass momentum, and (ii)~is considerably
strong, being governed by the same (large) in-plane dipole moment $d_\|$ as the optical transitions. For dark excitons, the exchange 
interaction is (i)~a short-range local field effect, so it weakly depends on momentum, (ii)~strong enough to produce a sizeable dark-
grey splitting $\delta\approx{0}.6\:\mbox{meV}\approx{7}\:\mbox{K}$ in WSe$_2$~\cite{Robert2017} which makes the dark component dominant
at low temperatures, and (iii)~at the same time weak enough to enable a control of the splitting and of the valley content by a perpendicular 
magnetic field $B_\perp$ of a few Tesla~\cite{Robert2017}. Importantly, by applying an in-plane magnetic field $B_\|$, one can transfer 
the in-plane oscillator strength from the higher-energy bright states to the dark-grey doublet~\cite{Molas2017, Zhang2017}, thus boosting 
the out-of-plane emission of light from both components of this doublet.

In this work we present an experimental implementation of the described ``dark exciton valley toolkit'' in a WSe$_2$ monolayer. Namely, 
we investigate the influence of the in-plane and/or out-of-plane magnetic fields on dark excitons emission. In particular, 
we experimentally demonstrate the coherent valley-superposition nature of the dark and grey excitons by transferring the oscillator strength 
from the bright excitons with an in-plane magnetic field and detecting a linearly polarized luminescence. We use a WSe$_2$ sample encapsulated 
in hexagonal boron nitride (hBN), which is known to significantly enhance the sample quality~\cite{WangMak2017,Cadiz2017,Ajai2017,Manca2017,Vaclavkova2018}.
In our experiment, the encapsulation leads to a very strong brightening effect, much more pronounced than observed earlier~\cite{Molas2017,Zhang2017}, 
and enables us to spectrally resolve the fine structure of the brightened dark excitons. We observed that the combination of 
the in-plane and out-of-plane magnetic fields leads to circularly-polarized emissions due to the dark and grey excitons.

The valley structure of intravalley excitons in a magnetic field of arbitrary orientation can be described using the model of Refs~\citenum{Slobodeniuk2016, Molas2017, Robert2017}.
Let us work in the basis of the four intravalley A-exciton states with zero center-of-mass momentum at zero magnetic field. These states
$|\mathbf{K}_\pm,s_c\rangle$ can be identified by the valley $\mathbf{K}_\pm$ and the conduction band spin projection $s_c=\uparrow,\downarrow$. The
valence band spin projection is fixed for the A~exciton: $s_v=\uparrow$ in $\mathbf{K}_+$ valley and $\downarrow$ in $\mathbf{K}_-$. The states 
$|\mathbf{K}_+,\uparrow\rangle$, $|\mathbf{K}_-,\downarrow\rangle$ are bright and are denoted by $|\mathbf{K}_\pm,\mbox{b}\rangle$, respectively. 
Their transition dipole matrix elements lie in the plane and correspond to the two circular polarizations, $\langle0|\hat{\mathbf{d}}|\mathbf{K}_\pm,\mathrm{b}\rangle=
-id_\|(\pm\mathbf{e}_x+{i}\mathbf{e}_y)$. The states $|\mathbf{K}_+,\downarrow\rangle$, $|\mathbf{K}_-,\uparrow\rangle$ are denoted as 
$|\mathbf{K}_\pm,\mbox{d}\rangle$ and have transition dipoles $\langle0|\hat{\mathbf{d}}|\mathbf{K}_\pm,\mathrm{d}\rangle=id_\perp\mathbf{e}_z$,
perpendicular to the plane~\cite{Glazov2014, Slobodeniuk2016,WangMarie2017, Robert2017, Slobodeniuk2019}, $d_\perp/d_\|\sim
0.01-0.1$~\cite{Slobodeniuk2016}.

In the basis $|\mathbf{K}_+,\mbox{b}\rangle$, $|\mathbf{K}_-,\mbox{b}\rangle$, $|\mathbf{K}_+,\mbox{d}\rangle$, $|\mathbf{K}_-,\mbox{d}\rangle$, 
the Hamiltonian matrix has the form
\begin{equation}
H=\left[\begin{array}{cccc}
\Delta+g_\mathrm{b}\mathcal{B}_z & 0 & g_\|\mathcal{B}_- & 0 \\
0 & \Delta-g_\mathrm{b}\mathcal{B}_z & 0 & g_\|\mathcal{B}_+ \\
g_\|\mathcal{B}_+ & 0 & \delta/2+g_\mathrm{d}\mathcal{B}_z & \delta/2 \\
0 & g_\|\mathcal{B}_- & \delta/2 & \delta/2-g_\mathrm{d}\mathcal{B}_z
\end{array}\right].\label{eq:Hamiltonian}
\end{equation}
Counting energies from the dark exciton state, we denoted by $\Delta\approx40\:\mbox{meV}$~\cite{Molas2017, Zhang2017, Robert2017, WangMarie2017} 
and $\delta\approx0.6\:\mbox{meV}$~\cite{Robert2017}, respectively, the dark-bright and dark-grey splitting at zero field. The parameter
$\delta$ characterizes the short-range exchange interaction which lifts the valley degeneracy of the dark excitons at $B_\perp=0$
and forms two new (grey and dark) eigenstates: $|\mathrm{G}\rangle, |\mathrm{D}\rangle=\big(|\mathbf{K}_+,\mathrm{d}\rangle \pm
|\mathbf{K}_-,\mathrm{d}\rangle\big)/\sqrt{2}$.

Other terms in equation~(\ref{eq:Hamiltonian}) describe the effect of the magnetic field, of the in-plane component,~$B_\|$, and of the
perpendicular one, $B_\perp$; we use the shorthand notation $\mathcal{B}_z\equiv\mu_\mathrm{B}B_\perp/2$, $\mathcal{B}_\pm\equiv\mu_\mathrm{B}(B_x\pm{i}B_y)/2$, 
where $\mu_\mathrm{B}=57.88\:\mu\mbox{eV/T}$ is the Bohr magneton. The perpendicular field~$B_\perp$ lifts the single-electron Kramers 
degeneracy between opposite spins in opposite valleys. This results in a valley-dependent shift of the excitonic levels with different 
Land\'e factors in the bright and dark-grey sectors. The in-plane magnetic field $\mathbf{B}_\parallel=(B_x,B_y)$ produces the Zeeman 
coupling ($g_\|\mu_\mathrm{B}B_\|$) of the electronic states with opposite spins in the same valley, and thus mixes the dark-grey and 
bright excitonic sectors, remaining diagonal in the valley index.

\begin{figure*}[t]
	\includegraphics[width=1.00\textwidth]{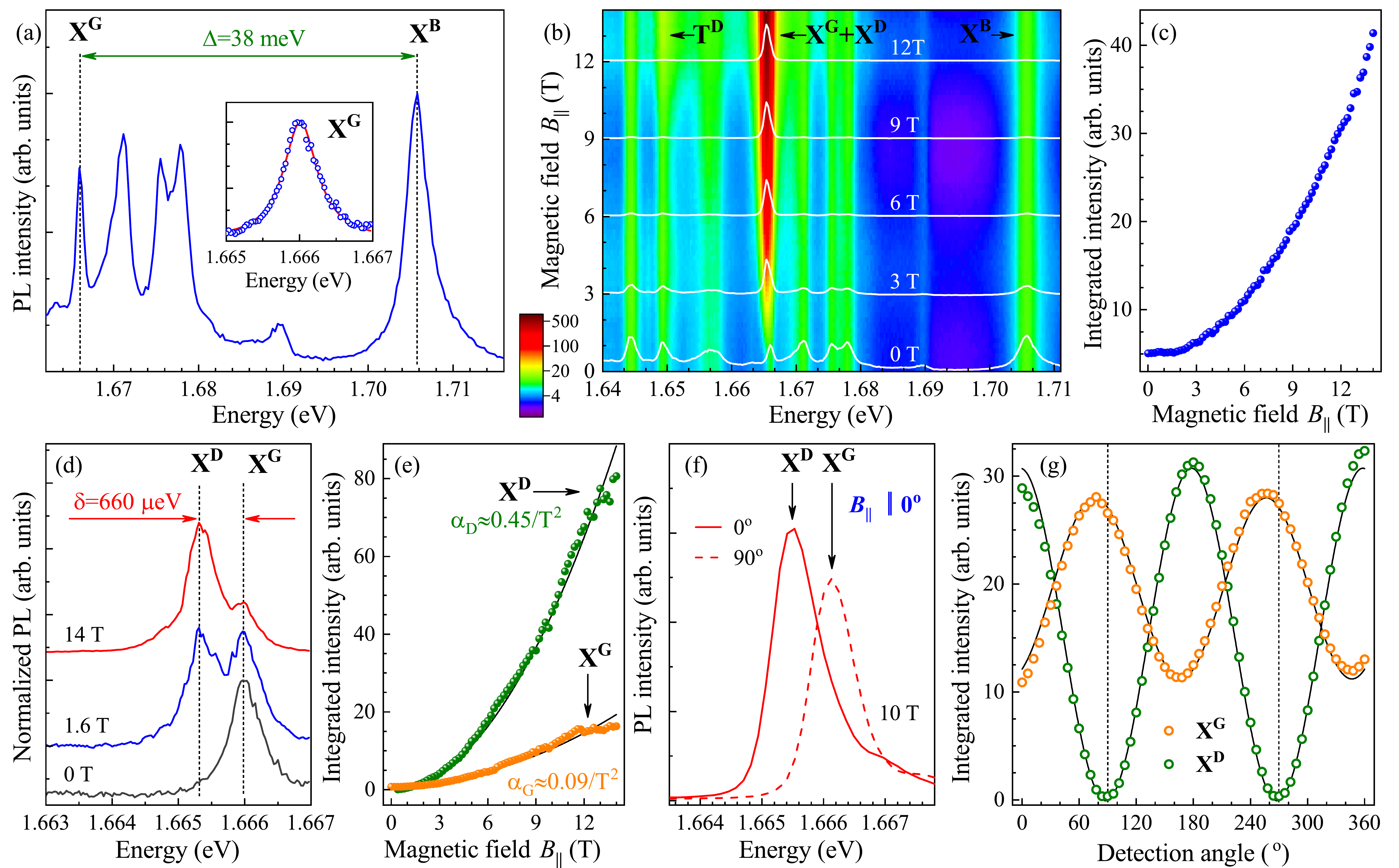}
	\caption{(a)~Low-temperature ($T$=4.2~K) PL spectrum measured on a WSe$_2$ monolayer encapsulated in hBN flakes, using an excitation energy 2.408~eV and a power of 50~$\mu$W, and zero magnetic field. The inset shows the high-resolution PL emission of the grey exciton. (b)~False-colour map of the PL response as a function of $B_\|$; white curves superimposed on the map represent the PL spectra normalized to the most intense peaks recorded at selected values of $B_\|$. (c)~Total emission as a function of $B_\|$ (frequency integral of panel~(b)). (d)~High resolution PL spectra normalized to the most intense peaks at selected values of $B_\|$. (e)~$B_\|$~dependence of the dark and grey exciton luminescence peaks' integrated intensities. The solid black curves represent quadratic fits.
		(f)~PL spectra at $B_\parallel=10~\mbox{T}$ recorded for two orthogonal linear polarizations oriented parallel (solid curve) and perpendicular (dashed curves) to the direction of $\mathbf{B}_\|$ (excitation energy 1.917~eV and power 50~$\mu$W). (g)~Polarization dependence of the dark and grey exciton luminescence peaks' integrated intensities measured at $B_\parallel$=10~T. } 
	\label{fig:brightening}
\end{figure*}

The magnetic field terms in Hamiltonian~(\ref{eq:Hamiltonian}) provide the experimental tools which enable one to control and probe 
the dark exciton states. At $B_\|=0$, one can tune continuously the valley content of the dark and grey excitons by changing~$B_\perp$.
Then, turning on the ~$B_\|$ field, one admixes the bright states to the dark-grey sector, thereby transferring the in-plane
oscillator strength; since $\Delta$ is the largest energy scale in Hamiltonian~(\ref{eq:Hamiltonian}), this admixture can be treated
perturbatively. Then, it turns out that the valley content of the eigenstates in the dark-grey sector can be probed by measuring linear
polarization of the light emitted now in the direction perpendicular to the monolayer plane. Consider an eigenstate $|\psi_\mathrm{d}\rangle$ 
at $B_\|=0$ in the form of a general linear combination, $|\psi_\mathrm{d}\rangle=\chi_+e^{i\phi_\mathrm{d}}|\mathbf{K}_+,\mathrm{d}\rangle+\chi_-e^{-i\phi_\mathrm{d}}|\mathbf{K}_-,\mathrm{d}\rangle$
with real $\chi_\pm$, $\chi_+^2+\chi_-^2=1$ and some phase $\phi_\mathrm{d}$. Let us apply $B_\|$ and detect the luminescence, emitted 
normally to the plane, with an analyzer selecting the electric field direction with the polar angle~$\phi_a$. Then, calculating perturbatively 
in $1/\Delta$ the corresponding in-plane dipole matrix element, we obtain the emitted light intensity:
\begin{eqnarray}\nonumber
&&I_\perp(\phi_a)\propto \left(d_\|\,\frac{g_\|\mu_\mathrm{B}B_\|}{\Delta}\right)^2
\times{}\\ &&\qquad{}\times
\left[(\chi_+-\chi_-)^2+4\chi_+\chi_-\sin^2(\phi_a+\phi_\mathrm{d}-\phi_B)\right],\label{eq:polarization}
\end{eqnarray}
where $\phi_B=\arctan(B_y/B_x)$ is the polar angle of the in-plane magnetic field. Below we present the experimental implementation of the described tools.

Figs~\ref{fig:brightening}(a)-(c) illustrate the results of micro-magneto-PL measurements performed in a wide spectral
range ($1.64 - 1.71$\,eV), covering the whole emission spectrum of our sample, but at the expense of rather low spectral resolution (0.8\,
nm). In contrast, the results shown in Figs~\ref{fig:brightening}(d),(e) and in the inset of Fig.~\ref{fig:brightening}(a)  refer to experiments 
with higher spectral resolution (0.1\,nm). All spectra were measured in the configuration of the normal (to the sample plane) coincidence
of the excitation and the collected beams (see Fig.~\ref{fig:excitons}(b) and Supplemental Material (SM) for details). The PL peaks, labelled 
as X$^\textrm{D}$ and X$^\textrm{G}$, are ascribed, correspondingly, to dark and grey states of the neutral exciton. Only X$^\textrm{G}$ is
observed in the spectra measured at zero magnetic field. We expect that the X$^\textrm{G}$-emission is directed predominantly along the 
monolayer plane, thought it is still visible in the spectra, due to a relatively high numerical aperture (NA) of the lenses used to collect 
the emitted light. Both components of the X$^\textrm{D}$/X$^\textrm{G}$ doublet contribute to the emission spectrum when the magnetic field
is applied (see, $e.g.$, Fig.~\ref{fig:brightening}(d)). Note that this doublet structure is not resolved in the spectra shown in Fig.~\ref{fig:brightening}(b), 
measured with low spectral resolution. These results are in good agreement with previous reports~\cite{Cadiz2017, Courtade2017, Robert2017, Barbone2018, Chen2018, Li2018,Zhou2017, WangMarie2017, Robert2017}: 
(i)~the grey exciton X$^\textrm{G}$-line is red-shifted by $38$~meV from the emission peak X$^\textrm{B}$ associated with the neutral bright 
exciton (see Fig.~\ref{fig:brightening}(a)), (ii)~the energy separation between the grey (X$^\textrm{G}$) and dark (X$^\textrm{D}$) excitons 
is $\delta=660\:\mu\mbox{eV}$ (see Fig.~\ref{fig:brightening}(d)), and (iii) the X$^\textrm{D}$ and X$^\textrm{G}$ lines are much narrower 
($\simeq0.6$~meV) than the X$^\textrm{B}$ line ($\simeq4$~meV) what indicates the significantly longer lifetimes of dark and grey
excitons as compared to the lifetime of the bright exciton (see SM for details).

Application of the in-plane magnetic field $B_\|$ leads to a strong brightening effect, which is remarkably pronounced in our encapsulated WSe$_2$ 
monolayer, much more than in previously probed WSe$_2$ monolayers on Si/SiO$_2$ substrates~\cite{Molas2017, Zhang2017}. This is demonstrated in 
Fig.~\ref{fig:brightening}(b) showing PL spectra measured in a wide spectral range, as a function of $B_\|$. With increasing $B_\|$, the intensity 
of the grey/dark exciton doublet (separate components not resolved) increases significantly, while the intensity of the bright exciton and of 
the lower energy features of the spectrum, stays practically at the same level. Note that the T$^\textrm{D}$ line observed at about 
1.65~eV in Fig.~\ref{fig:brightening}(b) also becomes significantly brighter at the largest magnetic fields ($B$$>$10~T). Recently, this peak was 
ascribed to the emission of the dark negatively charged exciton~\cite{Liu2019}. In Fig.~\ref{fig:brightening}(c), we plot the total emission 
intensity (integrated over the whole PL spectrum) as a function of~$B_\|$. At $B_\|$=14~T, the total emission intensity is enhanced by an order 
of magnitude and practically the whole luminescence of the sample is due to the dark/grey excitons doublet.

The PL response of our hBN/WSe$_2$/hBN sample, measured with better spectral resolution, enabled us, for the first time, to resolve the magnetic 
brightening of the dark and grey components separately. As can be seen in Fig.~\ref{fig:brightening}(d), while only the grey exciton is observed 
when $B_\|$=0, at fields above 1.6~T the X$^\textrm{D}$ line becomes more intense. The $B_\|$~dependence of the intensity of each component is 
expected to be $I=I_0+\alpha{B}_\|^2$ (Refs~\citenum{Molas2017,Zhang2017}), where the zero-field intensity $I_0$ vanishes for the dark component 
but is nonzero and depends on the NA of the excitation/detection lens for the grey component. The results of the quadratic fit are shown in 
Fig.~\ref{fig:brightening}(e). The $\alpha$ parameter turns out to be strongly different for the two components: $\alpha_\mathrm{D}=0.45\:\mbox{T}^{-2}$, while
$\alpha_\mathrm{G}=0.09\:\mbox{T}^{-2}$. This large difference is explained by the difference in the populations of the two states: at temperature 
$T$=4.2~K which corresponds to our experimental conditions, the population ratio is $e^{-\delta/k_BT}=0.16$, in reasonable agreement with the
measured ratio $\alpha_\mathrm{G}/\alpha_\mathrm{D}=0.2$.

Now, one can investigate the polarization properties of magnetically brightened dark and grey excitons, see Fig.~\ref{fig:brightening}(f).
We present in Fig.~\ref{fig:brightening}(g) the integrated PL intensity for each peak, as collected via the linear polarizer 
and measured as a function of the angle $\phi_a-\phi_B$ between the polarizer axis and the direction of the $\mathbf{B}_\|$ field. For the dark 
exciton, the polarization dependence fully agrees with equation~(\ref{eq:polarization}): indeed, at $B_\perp=0$, we have $\chi_+=-\chi_-=1/\sqrt{2}$ 
and $\phi_\mathrm{d}=0$, resulting in the transition dipole oriented along $\mathbf{B}_\|$. Our polarization measurement thus reveals experimentally 
the nature of the dark exciton state, which is a coherent superposition of the states in the two valleys.

\begin{figure}
	\includegraphics[width=0.5\textwidth]{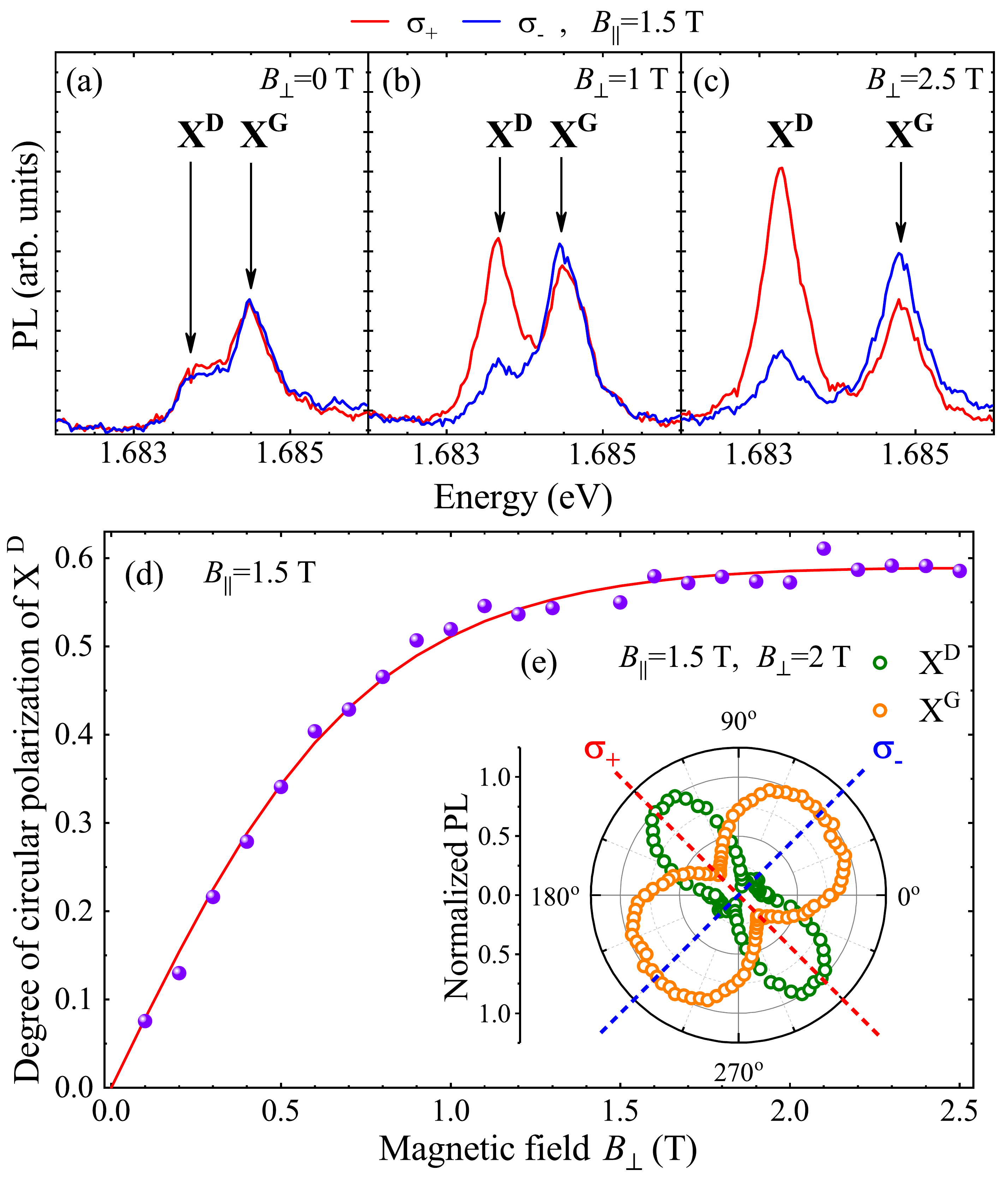}
	\caption{(a)-(c) PL spectra of dark and grey excitons in perpendicular magnetic field applied additionally to in-plane magnetic field of 1.5~T. (d) Emergence of the circular polarization of the dark exciton upon application of additional perpendicular magnetic field. Solid red line represents a fit of function described in SM. (e) PL intensity of X$^\textrm{D}$ and X$^\textrm{G}$ as a function of orientation of the $\lambda$/4 waveplate in detection. Settings 45$^\circ$ and 135$^\circ$ correspond to $\sigma_-$ and $\sigma_+$ polarization, respectively. The data were measured for orientations 0$^\circ$-180$^\circ$ and cloned to 180$^\circ$-360$^\circ$ range for the sake of presentation. Note that these results where measured on a different sample as those shown in Fig.~\ref{fig:brightening}.} 
	\label{fig:mixed}
\end{figure}

For the grey exciton, the measured polarization dependence corresponds to equation~(\ref{eq:polarization}) plus a constant background. The latter 
is due to large NA of the lens used to collect the emitted light. Indeed, equation~(\ref{eq:polarization}) describes emission perpendicular
to the monolayer and thus is valid in the limit of small NA. Our large-NA lens picks up the light emitted by the out-of-plane component $d_\perp$ 
of the transition dipole moment, which has no in-plane polarization. For the dark exciton at $B_\perp=0$, this component is absent.

Application of a magnetic field $B_\perp$ in the direction perpendicular to the monolayer plane shifts the energies in the two valleys in the opposite directions. 
Thus, the eigenstates in the dark-grey sector resulting from valley mixing by exchange coupling have no longer equal weights in the two valleys. 
As a result, both grey and dark exciton components have a finite out-of-plane optical dipole moment even at $B_\|$=0, and are visible in the 
luminescence spectra due to large NA of the lens (see SM for details). Note that the intensity of the dark exciton lines is independent of the detected helicity 
($\sigma_+$ or $\sigma_-$) of light.

In order to induce the $\sigma_\pm$ selectivity for the dark excitons emission, we propose to combine the effects of parallel ($B_\|$)
and perpendicular ($B_\perp$) magnetic fields. $B_\|$ gives rise to linear polarizations of the grey and dark excitons (see Fig.~\ref{fig:brightening}(f) 
and (g)), which are subsequently transformed by $B_\perp$ into elliptical polarizations. The experimental evidence corroborating this scenario is presented in 
Fig.~\ref{fig:mixed}(a)-(c). Upon increasing of the $B_\perp$ component of the magnetic field, both dark and grey excitons gain circular $\sigma_+$ and $\sigma_-$ 
polarizations, respectively. This allows us to determine that the $g$-factor of dark and grey excitons is negative as the one of the bright neutral exciton. 
The gradual increase in the circular degree of polarization, illustrated in Fig.~\ref{fig:mixed}(a)-(c), can be described in terms of the competition between 
the Zeeman shift $g_\mathrm{d}\mu_B B_\perp$ and the initial splitting $\delta$ and taking into account a finite NA of excitation/collection lens used in experiment.
The result of such competition is presented in Fig.~\ref{fig:mixed}(d) for the case of dark exciton emission, as an example. The blue dots (red curve) depicts the 
experimentally (theoretically) derived data of the circular polarization degree as a function of out-of-plane magnetic field $B_\perp$. Note that the value of the 
polarization degree saturates already at 60\% and not at 100\% as one can expect. The explanation of such phenomenon involves a finite contribution of the out-of-plane 
dipole components of X$^\textrm{D}$ and X$^\textrm{G}$ exciton states (see SM for details). To prove the dominant contribution of the circular polarization ($\sigma_\pm$)
for the dark and grey exciton emissions, they were measured as a function of orientation of the $\lambda$/4 waveplate in detection, see Fig.~\ref{fig:mixed}(f).

In this work, we investigated the PL response of a hBN-encapsulated WSe$_2$ monolayer subject to a magnetic field, whose direction lies in the monolayer plane, 
or perpendicular to it. Thanks to the encapsulation, the in-plane field leads to a very pronounced brightening effect on the dark lowest-energy excitons in WSe$_2$,
and most of the sample's luminescence comes from these dark excitons. Moreover, in our encapsulated sample one can resolve the fine structure of the magnetically 
brightened dark excitons, split by the exchange interaction which produces two coherent valley superpositions as eigenstates. We were able to probe the
coefficients of the superpositions by a polarization-resolved PL measurement. These coefficients can be changed by applying a perpendicular and/or parallel 
magnetic field. These findings open a perspective for controlled manipulation of the valley degree of freedom of long-lived dark excitons in monolayer S-TMDs.

The work has been supported by the ATOMOPTO project (TEAM programme of the Foundation for Polish Science, co-financed by the EU within the ERDFund), the EU 
Graphene Flagship project (no. 785219), the National Science Centre, Poland (grant no. \mbox{UMO-2018/31/B/ST3/02111}), the Nanofab facility of the Institut N\'eel, 
CNRS/UGA and the LNCMI-CNRS, a member of the European Magnetic Field Laboratory (EMFL). K.W. and T.T. acknowledge support 
from the Elemental Strategy Initiative conducted by the MEXT, Japan and and the CREST (JPMJCR15F3), JST.

\bibliographystyle{apsrev4-1}
\bibliography{Dark_WSe2}

\newpage
\onecolumngrid
\setcounter{figure}{0}
\setcounter{section}{0}
\renewcommand{\thefigure}{S\arabic{figure}}
\renewcommand{\thesection}{S\arabic{section}}

\begin{center}
	{\large{ {\bf Supplemental Material: \\ Probing and manipulating valley coherence of dark excitons in monolayer WSe$_2$}}}
	\vskip0.5\baselineskip{M. R. Molas,{$^{1,2}$} A. O. Slobodeniuk,{$^{1}$}, T.~Kazimierczuk,{$^{2}$} K. Nogajewski,{$^{1,2}$} M. Bartos,{$^{1}$}, P.~Kapu\'{s}ci\'{n}ski,{$^{1,3}$} K.~Oreszczuk,{$^{2}$} K. Watanabe,{$^{4}$} T. Taniguchi,{$^{4}$} C. Faugeras,{$^{1}$} D. M. Basko,{$^{5}$} and M. Potemski,{$^{1,2}$}}
	\vskip0.5\baselineskip{\em$^{1}$ Laboratoire National des Champs Magn\'etiques Intenses, CNRS-UGA-UPS-INSA-EMFL, 25, avenue des Martyrs, 38042 Grenoble, France \\$^{2}$ Faculty of Physics, University of Warsaw, ul. Pasteura 5, 02-093 Warszawa, Poland \\$^{3}$Department of Experimental Physics, Faculty of Fundamental Problems of Technology, Wroc{\l}aw University of Science and Technology, 27 Wybrze\.{z}e Wyspia\'{n}skiego, 50-370 Wroc{\l}aw, Poland \\$^{4}$National Institute for Materials Science, 1-1 Namiki, Tsukuba 305-0044, Japan \\$^{5}$Laboratoire de Physique et Mod\'elisation des Milieux Condens\'es, Universit\'e Grenoble Alpes and CNRS, 25 avenue des Martyrs, 38042 Grenoble, France}
\end{center}

This supplemental material provides: \ref{exp} description of preparation of the studied samples and used experimental setups, \ref{pl} low-temperature photoluminescence spectrum measured on the studied sample, \ref{lifetime} the influence of in-plane magnetic field on lifetimes of dark and grey excitons, \ref{faraday} the effect of perpendicular magnetic field on grey and dark excitons emissions, and \ref{polarization} polarization properties of dark excitons emission.

\section{S\lowercase{amples and experimental setups}\label{exp}}

The active part of our sample consists of a monolayer of WSe$_2$, which has been encapsulated in hexagonal boron nitride (hBN) and deposited on a bare Si
substrate. It was fabricated by two-stage polydimethylsiloxane (PDMS)-based~\cite{Gomez2014} mechanical exfoliation of WSe$_2$
and hBN bulk crystals. Micro-magneto-photoluminescence (PL) measurements (spatial resolution $\sim$5~$\mu$m) were carried out at liquid helium
temperature using two different experimental setups. In each setup, the sample was placed on top of a $x-y-z$ piezo-stage kept
at $T$=4.2~K and was excited using laser diodes with either 515~nm  wavelength (2.408~eV photon energy) or 647~nm
wavelength (1.917~eV photon energy). The emitted light was dispersed with a 0.5~m long monochromator and detected with a
charge coupled device (CCD) camera.  The data presented in Figs~2(a)-(e) and~\ref{fig:Bperp} refer to measurements performed
with the fiber-optic based setup and were carried out in magnetic fields up to 14\,T. Unpolarized emission was measured in Voigt 
configuration (see Fig.~2(a)-(e)), whereas the $\sigma_-$ and $\sigma_+$-polarized components of the emitted light were resolved 
in experiments (see Fig.~\ref{fig:Bperp}) performed in the Faraday configuration (polarizers placed in a close vicinity of the 
sample, in between the sample and optical fibers). The data presented in Figs~2(f)-(g) refer to experiments carried out 
using a split-coil (10~T) superconducting magnet with a free-beam-optics arrangement. Fig.~3 illustrates
results obtained with the aid of a system of two split-coil superconducting magnets with a free-beam-optics arrangement allowing 
to apply independently in-plane and out-of-plane magnetic fields. The linear and circular polarizations of the emissions were 
analyzed using a set of polarizers and waveplates ($\lambda$/2 and $\lambda$/4) placed directly in front of the spectrometers. 
The data presented in Figs~2, ~3, ~\ref{fig:pl} and ~\ref{fig:Bperp} refer to spectra measured with different spectral resolutions:
0.8~nm --- Figs~2(a)-(c), ~\ref{fig:pl} and ~\ref{fig:Bperp}; 0.2\,nm --- Fig.~2(f)-(g); 0.1~nm --- the inset of Fig.~2(a), Fig.~2(d)-(e) and Fig.~3.

\section{L\lowercase{ow-temperature} PL\lowercase{ spectrum of the }WS\lowercase{e$_2$ monolayer}\label{pl}}

\begin{figure}[t]
	\includegraphics[width=0.5\textwidth]{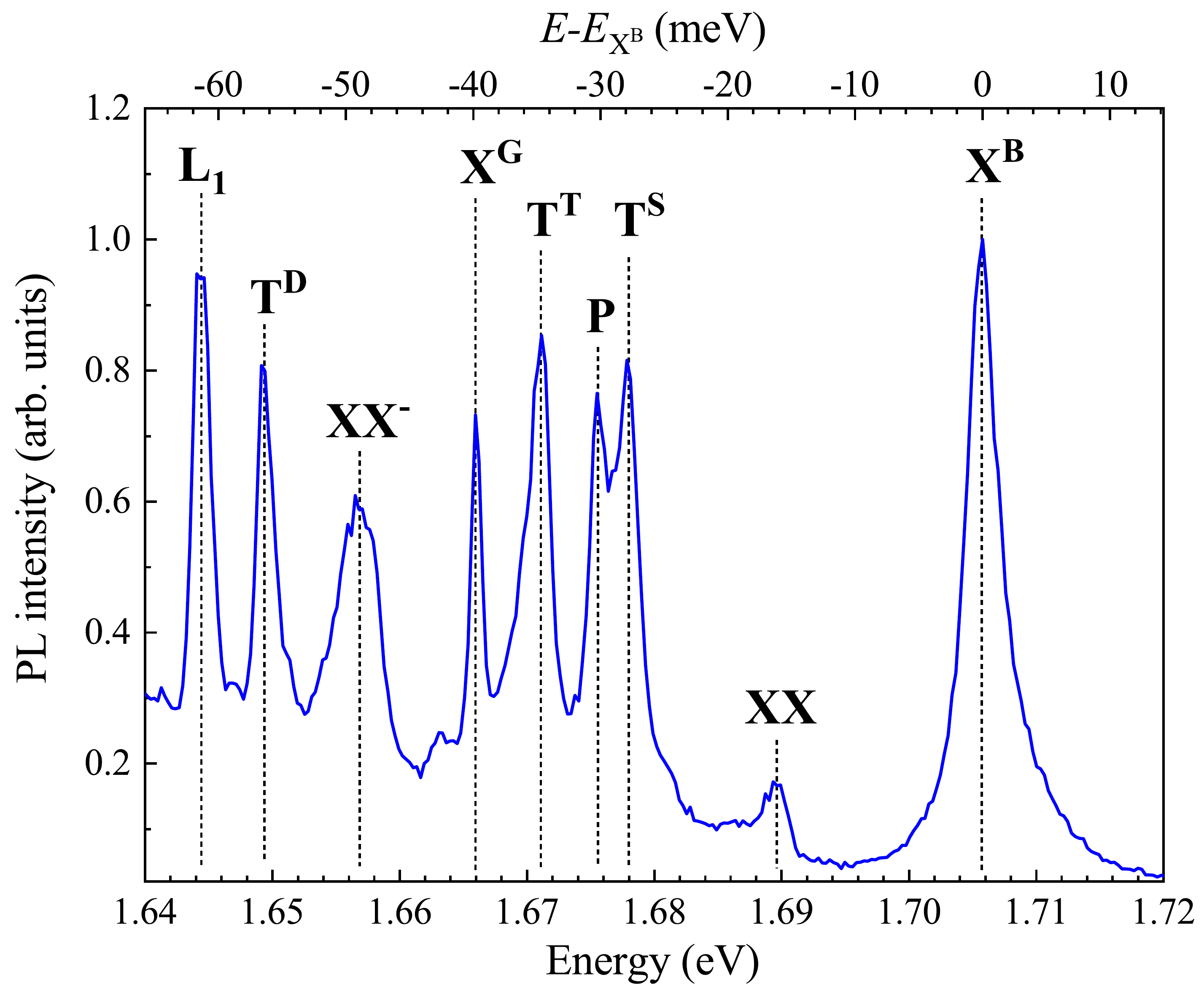}
	\caption{Low-temperature ($T$=4.2~K) PL spectrum measured on a WSe$_2$ monolayer encapsulated in hBN flakes, using an excitation energy 2.408~eV and a power of 50~$\mu$W, and zero magnetic field.} 
	\label{fig:pl}
\end{figure}

Fig.~\ref{fig:pl} demonstrates the photoluminescence spectrum measured on a WSe$_2$ monolayer encapsulated in hBN flakes.
The PL spectrum displays several emission lines with a similar characteristic pattern already reported in a number of previous works on WSe$_2$
monolayers embedded in between hBN flakes~\cite{Courtade2017, Li2018, Chen2018, Barbone2018, Paur2019, Liu2019}. In accordance with these reports, the assignment
of the observed emission lines is as follows: X$^\textrm{B}$ --- a bright neutral exciton formed in the vicinity of the A exciton; 
XX --- a neutral biexciton; T$^\textrm{S}$ and T$^\textrm{T}$ --- singlet (intravalley) and triplet (intervalley) negatively charged excitons,
respectively; X$^\textrm{G}$ --- a grey exciton; XX$^\textrm{-}$ --- a negatvely charged biexciton; T$^\textrm{D}$ - a negatively charged dark exciton;
L$_\textrm{1}$ --- a so-called localized exciton. The attribution of the P line is not clear, as in the most reported PL spectra this 
peak was absent~\cite{Li2018, Chen2018, Barbone2018, Paur2019, Liu2019}, but was observed in the neutral regime (between $n$ and $p$-type 
doping) in Ref.~\citenum{Courtade2017}. As the energy separation between X$^\textrm{B}$ and P lines of about 31 meV is close to the longitudinal optical 
(LO) phonon energy ($E_\textrm{LO}$=32~meV~\cite{Zhang2015}), we ascribed tentatively this peak to the LO phonon replica of the bright neutral exciton.

\newpage
\section{T\lowercase{he influence of in-plane magnetic field on lifetimes of grey and dark excitons}\label{lifetime}}

\begin{figure}[h]
	\includegraphics[width=0.5\textwidth]{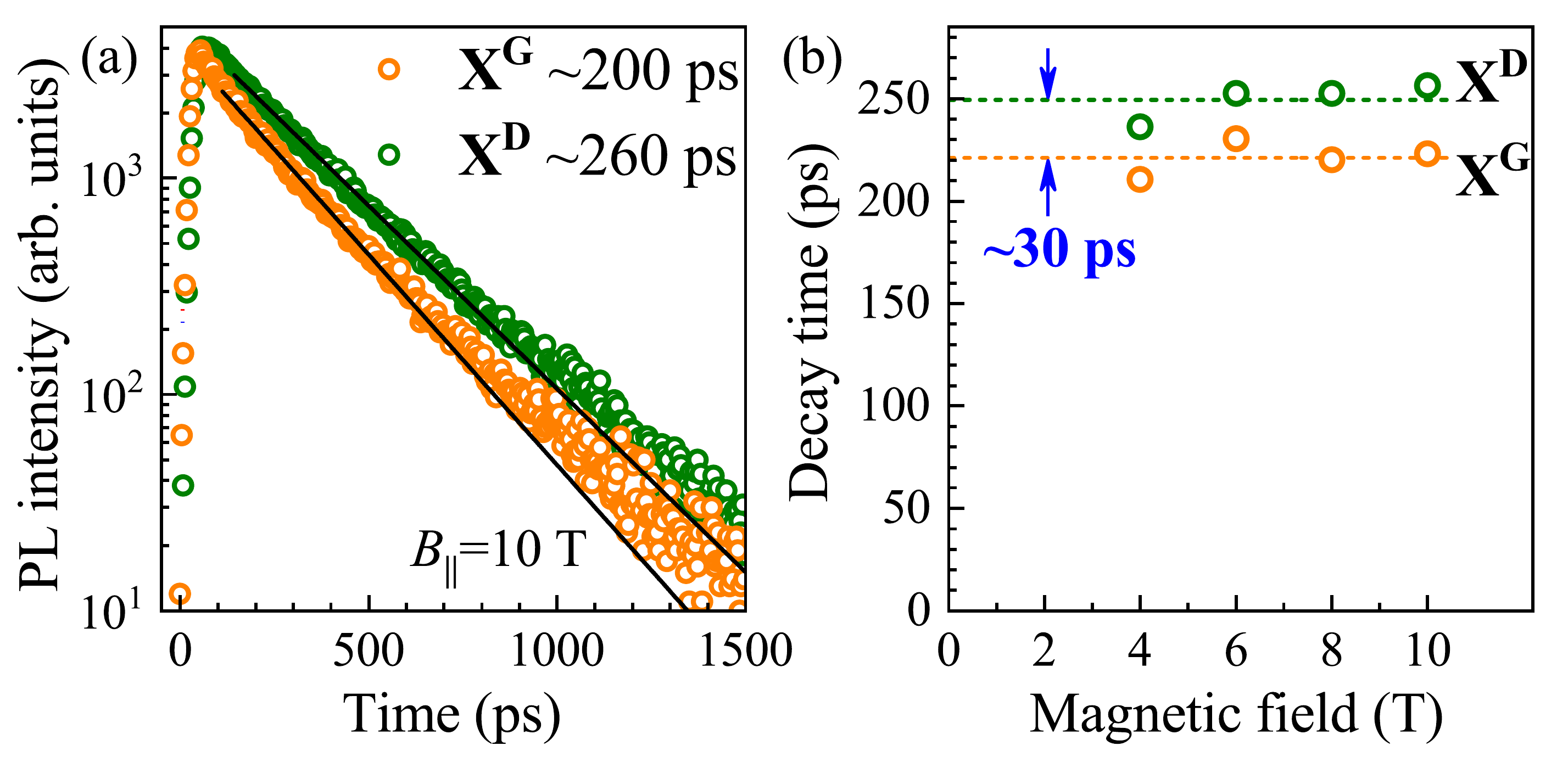}
	\caption{(a)~Low-temperature ($T$=4.2~K) time resolved PL measured on a WSe$_2$ monolayer encapsulated in hBN flakes under in-plane magnetic field $B_\|$=10~T, (b) The obtained decay times of the X$^\textrm{G}$ and  X$^\textrm{D}$ lines for a few magnetic fields.} 
	\label{fig:lifetime}
\end{figure}

To get more information on the grey and dark excitons, we measured their lifetimes by time-resolved photoluminescence
spectroscopy, when an in-plane magnetic field is applied. Fig.~\ref{fig:lifetime}(a) shows the X$^\textrm{G}$ and  X$^\textrm{D}$
emissions kinetics following a pulsed laser excitation, while Fig.~\ref{fig:lifetime}(b) summarizes their obtained decay times 
for a few values of magnetic fields. We found that the grey and dark excitons lifetimes are on the order of hundreds ps and are
significantly longer as compared to the reported $\sim$2 ps lifetime of the bright neutral exciton in WSe$_2$ monolayer~\cite{Robert2016,Robert2017}.  
As can be appreciated in Fig.~\ref{lifetime}(b), we do not observe an effect of an in-plane magnetic field on the X$^\textrm{G}$ 
and  X$^\textrm{D}$ lifetimes. Consequently, the extrapolated values of the grey and dark excitons at zero magnetic field are of 
about 220~ps and 250~ps, respectively. We note also that the obtained X$^\textrm{G}$ lifetime is almost two time larger than 
the reported previously in Ref.~\citenum{Robert2017}.

\section{T\lowercase{he effect of perpendicular magnetic field on grey and dark excitons emissions}\label{faraday}}

\begin{figure}
	\includegraphics[width=0.5\linewidth]{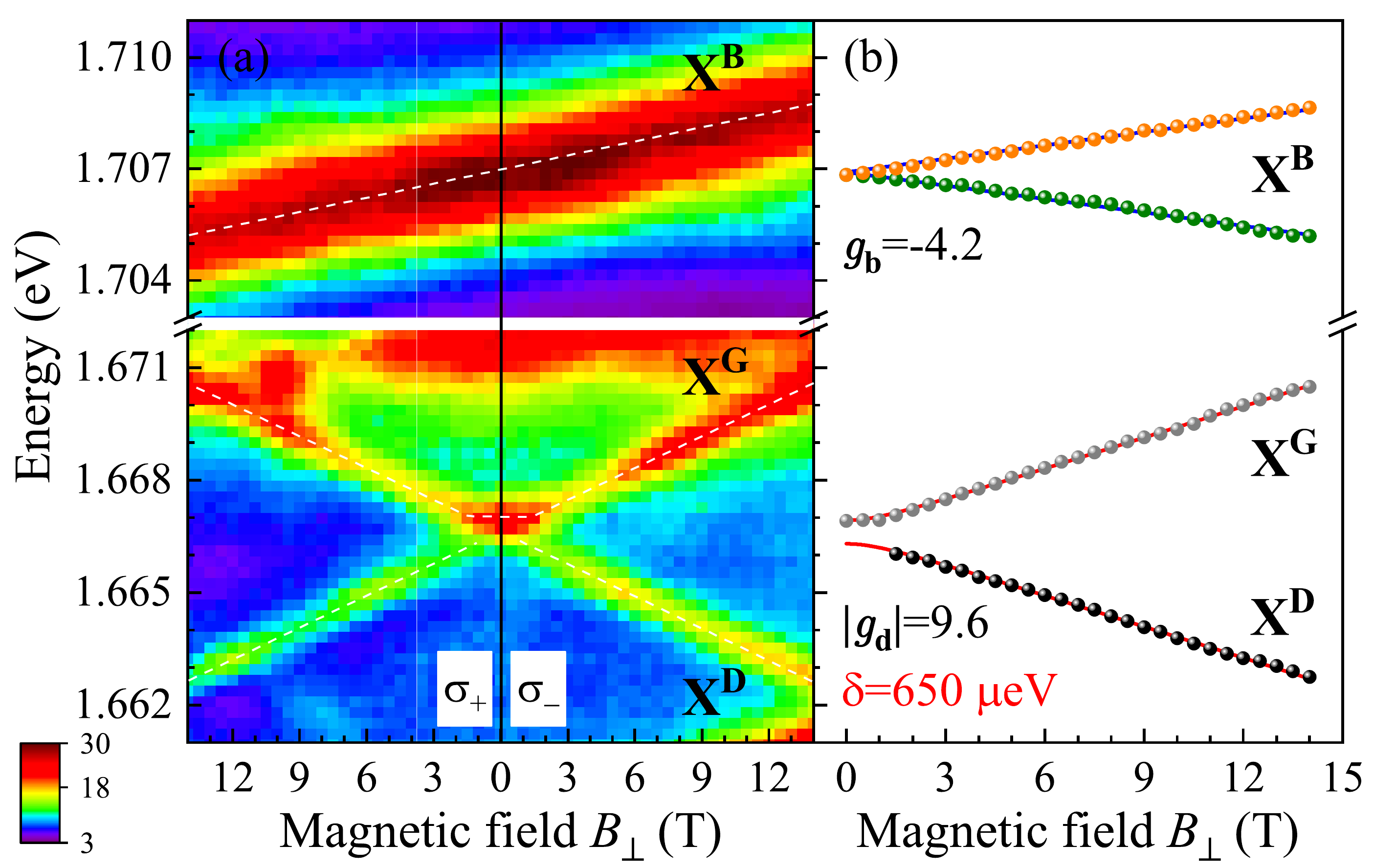}
	\caption{
		(a) False-colour map of the helicity-resolved ($\sigma_\pm$) PL spectra as a function of $B_\perp$ (excitation energy 2.408~eV and power 50 $\mu$W). The
		dashed white lines are guides to the eye. (b) Magnetic field dependence of the dark and bright exciton energies. The green and orange points correspond 
		to the $\sigma_+$- and $\sigma_-$-polarized components of the bright exciton resonance, respectively. The grey and black points denote correspondingly 
		the averaged energies of the grey and dark excitons for both circular polarizations. The solid blue and red curves represent fits with $E_{\mathrm{b},\pm}=E_\mathrm{B}\pm{g}_\mathrm{b}\mu_\mathrm{B}B_\perp/2$
		and equation~(\ref{eq:EGD}), respectively.} 
	\label{fig:Bperp}
\end{figure}

The magneto-luminescence spectra for two opposite circular polarizations, measured in the Faraday configuration are shown in Fig.~\ref{fig:Bperp}\,a. The
bright exciton emission splits into two circularly polarized components centered at energies $E_{\mathrm{b},\pm}=E_\mathrm{B}\pm{g}_\mathrm{b}\mu_\mathrm{B}B_\perp/2$,
due to the valley Zeeman effect, as can be appreciated at the top of Fig.~\ref{fig:Bperp}a,b. We obtain $g_\mathrm{b}=-4.2$, in agreement with previously 
reported values~\cite{Srivastava2015, Aivazian2015, WangGlazov2015, Mitioglu2015, Robert2017, Koperski2019}.

The magnetic field evolution of the dark and grey exciton peaks is different. First, the intensity of dark(grey) exciton line is the same in both circular
polarizations. Secondly, the peak energies are separated by the the exchange gap~$\delta$:
\begin{equation}
E_\mathrm{G,D}=E_\mathrm{d}\pm \frac{1}{2}\sqrt{\delta^2+(g_\mathrm{d} \mu_\mathrm{B} B_\perp)^2}.
\label{eq:EGD}
\end{equation}
From the fit we extract $\delta=650\:\mu\mbox{eV}$ and $|g_\mathrm{d}|=9.6$, in agreement with Ref.~\citenum{Robert2017} ($\delta=600\:\mu\mbox{eV}$ and
$|g_\mathrm{d}|=9.4$) and our measurements in the in-plane field, described above ($\delta=660\:\mu\mbox{eV}$). In the single particle picture, the Zeeman shift 
in a perpendicular magnetic field has two contributions: the spin and the valley-orbital~\cite{Aivazian2015, Srivastava2015, Li2014, Macneill2015, Koperski2017,Koperski2019}. 
The latter is the same for bright and dark excitons from the same valley. Hence, from the difference in the $g$~factors of dark and bright excitons we can extract 
the spin $g$~factor in the conduction band, which amounts to~2.7. Note that due to the results performed in combination of the in-plane and 
out-of-plane magnetic fields, we were able to determine that the grey-dark $g$-factor is negative similarly as for the bright neutral exciton and equals $g_\mathrm{d}=-9.4$.

\section{P\lowercase{olarization properties of dark excitons emission}\label{polarization}}

We consider the lowest energy (dark) exciton states of S-TMD monolayer in the presence of non-zero out-of-plane $B_\perp$ and in-plane $\mathbf{B}_\parallel$ magnetic fields.
$B_\perp$ couples $|\mathbf{K}_+,\mathrm{d}\rangle$ and $|\mathbf{K}_-,\mathrm{d}\rangle$ dark states  of opposite valleys. $\mathbf{B}_\parallel$ mixes bright $|\mathbf{K}_\pm,\mathrm{b}\rangle$ and dark $|\mathbf{K}_\pm,\mathrm{d}\rangle$ states within the same valley. Taking into account both effects and considering the bright-dark exciton coupling perturbatively in the parameter $g_\parallel\mathcal{B}_\parallel/\Delta\equiv g_\parallel\mu_B|\mathbf{B}_\parallel|/2\Delta\ll1$, one derives the lowest energy exciton states
\begin{align}
\label{eq:dh}
&|D_h\rangle=\mathcal{N}
\Big[-\frac{g_\parallel\mathcal{B}_\parallel}{\Delta}e^{-i\phi_B}
\sin\Big(\frac{\vartheta}{2}+\frac{\pi}{4}\Big),\,  -\frac{g_\parallel\mathcal{B}_\parallel}{\Delta}e^{i\phi_B}
\cos\Big(\frac{\vartheta}{2}+\frac{\pi}{4}\Big),\,
\sin\Big(\frac{\vartheta}{2}+\frac{\pi}{4}\Big),\,
\cos\Big(\frac{\vartheta}{2}+\frac{\pi}{4}\Big)\Big],\\
\label{eq:lh}
&|D_l\rangle=\mathcal{N}
\Big[-\frac{g_\parallel\mathcal{B}_\parallel}{\Delta}e^{-i\phi_B}
\cos\Big(\frac{\vartheta}{2}+\frac{\pi}{4}\Big),\,  \frac{g_\parallel\mathcal{B}_\parallel}{\Delta}e^{i\phi_B}
\sin\Big(\frac{\vartheta}{2}+\frac{\pi}{4}\Big),\,
\cos\Big(\frac{\vartheta}{2}+\frac{\pi}{4}\Big),\,
-\sin\Big(\frac{\vartheta}{2}+\frac{\pi}{4}\Big)\Big],
\end{align}
written in the basis $\{|\mathbf{K}_+,\mathrm{b}\rangle,|\mathbf{K}_-,\mathrm{b}\rangle,|\mathbf{K}_+,\mathrm{d}\rangle,
|\mathbf{K}_-,\mathrm{d}\rangle\}$. Here $\vartheta=\arcsin\Big[g_\mathrm{d}\mu_BB_\perp/\sqrt{\delta^2+\big(g_\mathrm{d}\mu_BB_\perp\big)^2}\Big]
\in[-\pi/2,\pi/2]$ and $\mathcal{N}=1/\sqrt{1+\big(g_\parallel\mathcal{B}_\parallel/\Delta\big)^2}$. The indices $h$ and $l$ represents the higher and lower energy branches of these new exciton states, with corresponding
energies up to quadratic in $g_\parallel\mathcal{B}_\parallel/\Delta$ terms
\begin{equation}
\left(\begin{array}{c}
E_h \\
E_l
\end{array}\right)=
\frac{\delta}{2}\pm\frac12\sqrt{\delta^2+\big(g_\mathrm{d}\mu_BB_\perp\big)^2}-
\frac{\big(g_\parallel\mu_B\mathbf{B}_\parallel\big)^2}{4\Delta}.
\end{equation}
Note that $|D_h\rangle$ and $|D_l\rangle$ states turn into gray $|G\rangle$ and dark $|D\rangle$ ones respectively  in the limit of the zero magnetic field. Therefore, one
can call $|D_h\rangle$ and $|D_l\rangle$ also dark exciton states for brevity.

The new dark excitons are optically active and their radiation patterns can be determined from the corresponding transition dipole matrix elements
\begin{align}
&\langle0|\mathbf{d}|D_h\rangle= i\mathcal{N}d_\parallel\frac{g_\parallel\mathcal{B}_\parallel}{\Delta}\Big\{e^{-i\phi_B}
\sin\Big(\frac{\vartheta}{2}+\frac{\pi}{4}\Big)[\mathbf{e}_x+i\mathbf{e}_y] +
e^{i\phi_B}\cos\Big(\frac{\vartheta}{2}+\frac{\pi}{4}\Big)[-\mathbf{e}_x+i\mathbf{e}_y]\Big\}+
i\mathcal{N}\sqrt{2}d_\perp\cos\Big(\frac{\vartheta}{2}\Big)\mathbf{e}_z ,\\
&\langle0|\mathbf{d}|D_l\rangle=i\mathcal{N}d_\parallel \frac{g_\parallel\mathcal{B}_\parallel}{\Delta}\Big\{e^{-i\phi_B}
\cos\Big(\frac{\vartheta}{2}+\frac{\pi}{4}\Big)[\mathbf{e}_x+i\mathbf{e}_y] -
e^{i\phi_B}\sin\Big(\frac{\vartheta}{2}+\frac{\pi}{4}\Big)[-\mathbf{e}_x+i\mathbf{e}_y]\Big\} -i\mathcal{N}\sqrt{2}d_\perp\sin\Big(\frac{\vartheta}{2}\Big)\mathbf{e}_z.
\end{align}
Both dipole vectors have the structure $\mathbf{d}_\alpha\equiv\langle0|\mathbf{d}|D_\alpha\rangle=a_\alpha[\mathbf{e}_x+i\mathbf{e}_y]+ b_\alpha[-\mathbf{e}_x+i\mathbf{e}_y]+c_\alpha\mathbf{e}_z,\,\alpha=h,l$ and can be considered simultaneously.

\begin{figure*}[t]
	\centering
	\includegraphics[width=0.55\linewidth]{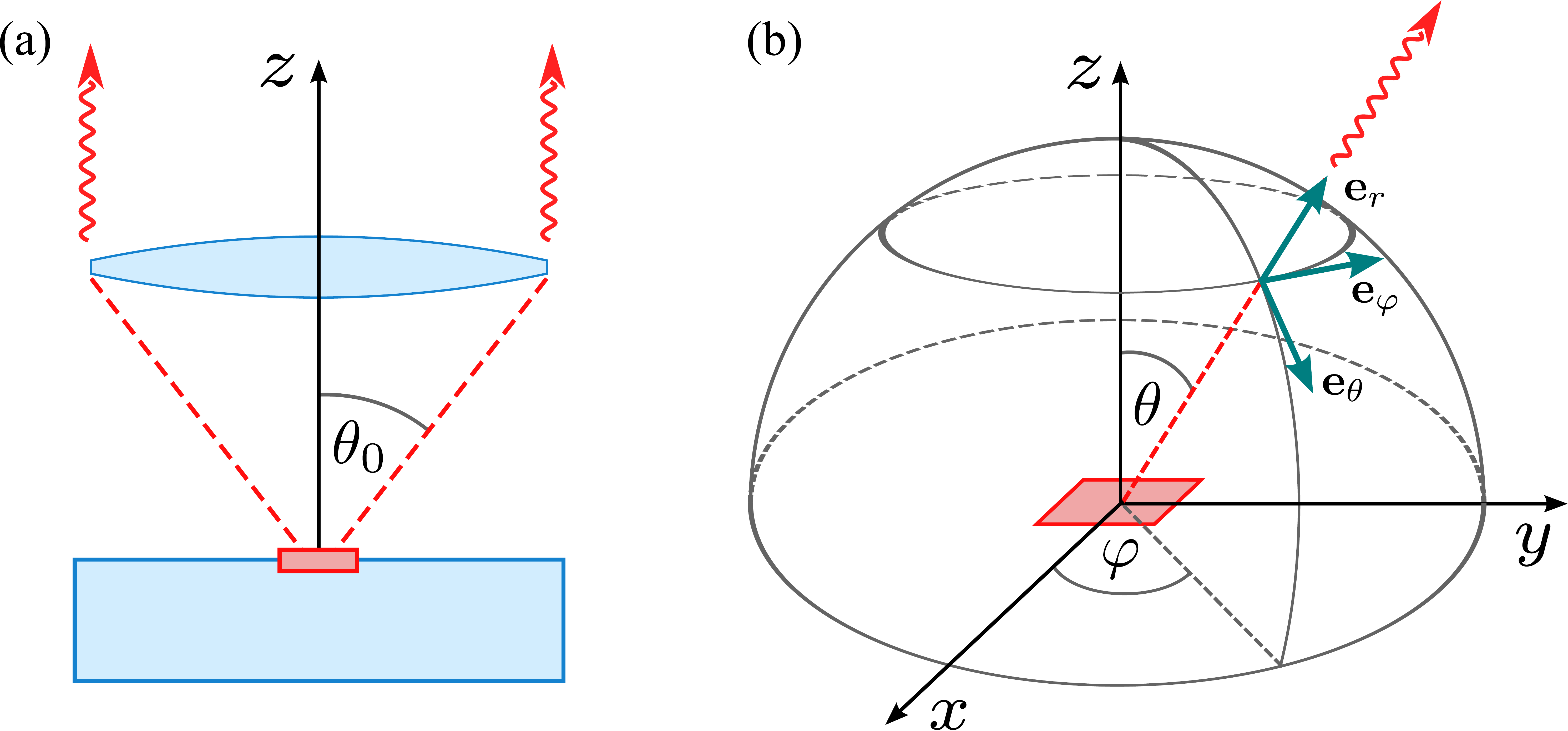}%
	\caption{(a) The side view of the main part of the setup. Red dashed lines represent the solid angle (with colatitude $\theta_0$), all the photons inside which are absorbed by
		the lens. The wavy arrows represent the photons outcoming from the lens. The red rectangular in the focus of the lens depicts S-TMD sample placed on Si/SiO$_2$ substrate (blue rectangular). (b) The schematic view of the trace of photons (dashed red line), emitted from the sample (red parallelogram) in the $(\theta, \varphi)$ direction. Three blue-green arrows represent the spherical basis vectors.}
	\label{fig:fig_1}
\end{figure*}

The polarization degree of dark exciton emission is defined as
\begin{equation}
P_\alpha=\frac{I_\alpha(\sigma_+)-I_\alpha(\sigma_-)}{I_\alpha(\sigma_+)+I_\alpha(\sigma_-)}.
\end{equation}
Here $I_\alpha(\sigma_\pm)$ is the total intensity of the $\sigma_\pm$ polarized light passing through the lens (see Fig.~\ref{fig:fig_1}(a)). $I_\alpha(\sigma_\pm)$ can be estimated by calculating the number of $\sigma_\pm$ polarized photons propagating inside the solid angle, defined by the colatitude $\theta_0$. We assume that the sample is placed in the focus of the lens, its size is much smaller than the size of the lens, and that the circular polarization of a photon is not affected as it passes through the lens.

The amount of $\sigma_\pm$ polarized photons, which are emitted from the sample, depends on the direction of their propagation. This direction is defined by spherical angles
$(\theta,\varphi)$ or radial unit vector $\mathbf{e}_r$ (see Fig.~\ref{fig:fig_1}(b)). The corresponding total intensities of light with fixed polarization can be evaluated as the integral
\begin{equation}
\label{eq:intensity}
I_\alpha(\sigma_\pm)\propto \int_0^{2\pi}d\varphi \int_0^{\theta_0} \sin\theta\,d\theta\, |\mathbf{d}_\alpha\cdot\mathbf{e}_\pm(\theta,\varphi)|^2.
\end{equation}
Here $\mathbf{e}_\pm(\theta,\varphi)$ are the polarization vectors of the $\sigma_\pm$ polarized photons propagating in the direction defined by the angles $(\theta,\varphi)$.
In order to write $\mathbf{e}_\pm$ explicitly, we introduce the local orthogonal unit vectors in the directions of increasing $r$, $\theta$ and $\varphi$, respectively
(see Fig.\ref{fig:fig_1}~b)
\begin{align}
&\mathbf{e}_r=\sin\theta\cos\varphi\,\mathbf{e}_x+\sin\theta\sin\varphi\,\mathbf{e}_y+\cos\theta\,\mathbf{e}_z,\\
&\mathbf{e}_\theta=\cos\theta\cos\varphi\,\mathbf{e}_x+\cos\theta\sin\varphi\,\mathbf{e}_y-\sin\theta\,\mathbf{e}_z,\\
&\mathbf{e}_\varphi=-\sin\varphi\,\mathbf{e}_x+\cos\varphi\,\mathbf{e}_y.
\end{align}
The spherical basis $\{\mathbf{e}_\theta,\mathbf{e}_\varphi,\mathbf{e}_r\}$ has the same relative orientation as $\{\mathbf{e}_x,\mathbf{e}_y,\mathbf{e}_z\}$.
Hence, it is natural to define $\mathbf{e}_\pm(\theta,\varphi)\equiv(\pm\mathbf{e}_\theta+i\mathbf{e}_\varphi)/\sqrt{2}$.
Taking the corresponding scalar products 
\begin{align}
&\mathbf{d}_\alpha\cdot\mathbf{e}_\pm(\theta,\varphi)=-\frac{a_\alpha}{\sqrt{2}}(1\mp\cos\theta)e^{i\varphi}-
\frac{b_\alpha}{\sqrt{2}}(1\pm\cos\theta)e^{-i\varphi}\mp \frac{c_\alpha}{\sqrt{2}}\sin\theta,
\end{align}
and then using (\ref{eq:intensity}) we obtain
\begin{equation}
P_\alpha=\frac{6(|b_\alpha|^2-|a_\alpha|^2)\cos^2\big(\frac{\theta_0}{2}\big)}{(|b_\alpha|^2+|a_\alpha|^2)
	(4+\cos\theta_0+\cos^2\theta_0)+2|c_\alpha|^2\sin^2\big(\frac{\theta_0}{2}\big)(2+\cos\theta_0)}.
\end{equation}
Finally, the expressions for polarization degrees of higher-energy and lower-energy states read
\begin{align}
&P_h=\frac{-6\cos^2\big(\frac{\theta_0}{2}\big)\,\sin\vartheta}{
	4+\cos\theta_0+\cos^2\theta_0+\Big(\frac{4\Delta}{g_\parallel\mu_BB_\parallel}\frac{d_\perp}{d_\parallel}
	\Big)^2\sin^2\big(\frac{\theta_0}{2}\big)(2+\cos\theta_0)
	\,\cos^2\big(\frac{\vartheta}{2}\big)},\\
&P_l=\frac{6\cos^2\big(\frac{\theta_0}{2}\big)\,\sin\vartheta}{
	4+\cos\theta_0+\cos^2\theta_0+\Big(\frac{4\Delta}{g_\parallel\mu_BB_\parallel}\frac{d_\perp}{d_\parallel}
	\Big)^2\sin^2\big(\frac{\theta_0}{2}\big)(2+\cos\theta_0)
	\,\sin^2\big(\frac{\vartheta}{2}\big)}.
\end{align}

\begin{figure*}[t]
	\centering
	\includegraphics[width=0.4\linewidth]{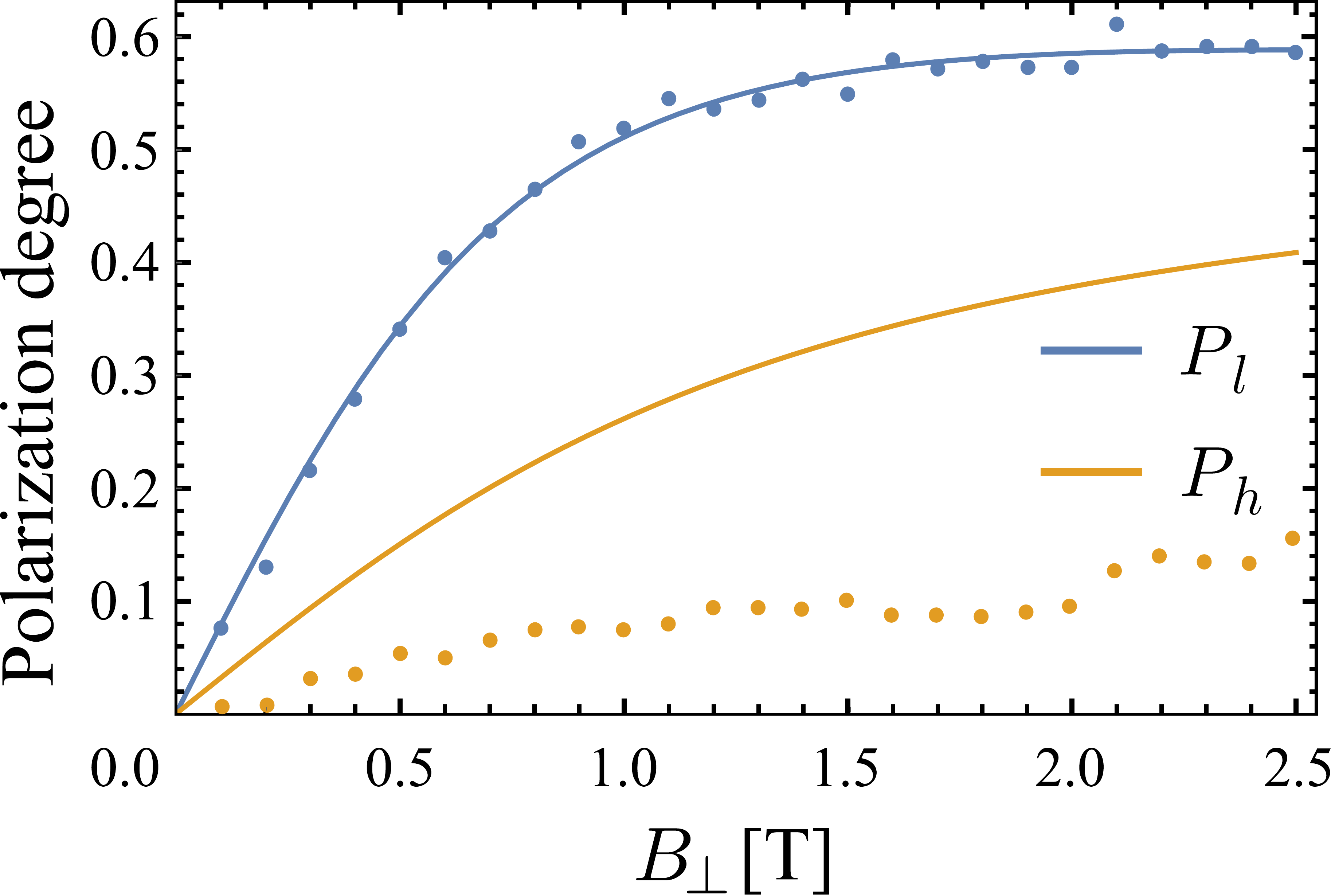}%
	\caption{Absolute values of the polarization degrees of lower ($P_l$) and higher-energy ($P_h$) dark excitons as a fucnction of the out-of-plane magnetic field $B_\perp$.
		Blue and yellow dots depict the corresponding experimental data.
		Blue line represents the best fit for lower-energy dark excitons, with parameters $\theta_0=1.139\approx 65^\circ$ and $d_\perp/d_\parallel=0.0035$. Yellow line corresponds
		to $P_h$ curve with the same parameters.}
	\label{fig:fig_2}
\end{figure*}

We fit the experimental data for lower-energy dark excitons polarization degree using the following parameters: $\Delta=38$ meV, $\mu_B=0.05788$ meV/T, $B_\parallel=|\mathbf{B}_\parallel|=1.5$ T, $|g_\mathrm{d}|=9.6$, $g_\parallel=2$, $\delta=0.65$ meV. The best fit corresponds to $\theta_0=1.139\approx 65^\circ$ and $d_\perp/d_\parallel=0.0035$, and is represented by a blue curve in Fig.~\ref{fig:fig_2}. Using the same parameters we plot the curve for polarization degree for higher-energy dark excitons, see an orange curve in Fig.~\ref{fig:fig_2}. For the considered range of parameters, the second curve approximates badly the corresponding experimental points. Probably, in the latter case there are additional unaccounted mechanisms, which reduce strongly the polarization degree of higher-energy exciton states.

\end{document}